# Lord Kelvin's atmospheric electricity measurements

K.L. Aplin(1) and R.G. Harrison(2)

(1) Physics Department, University of Oxford, Denys Wilkinson Building, Keble Road, Oxford OX1 3RH
(2) Department of Meteorology, University of Reading, PO Box 243, Earley Gate, Reading RG6 6BB

**Abstract**

Lord Kelvin (William Thomson) made important contributions to the study of atmospheric electricity during a brief but productive period from 1859-1861. By 1859 Kelvin had recognised the need for "incessant recording" of atmospheric electrical parameters, and responded by inventing both the water-dropper equaliser for measuring the atmospheric Potential Gradient (PG), and photographic data logging. The water-dropper equaliser was widely adopted internationally and is still in use today. Following theoretical considerations of electric field distortion by local topography, Kelvin developed a portable electrometer, using it to investigate the PG on the Scottish island of Arran. During these environmental measurements, Kelvin may have unwittingly detected atmospheric PG changes during solar activity in August/September 1859 associated with the "Carrington event", which is interesting in the context of his later statements that solar magnetic influence on the Earth was impossible. Kelvin's atmospheric electricity work presents an early representative study in quantitative environmental physics, through the application of mathematical principles to an environmental problem, the design and construction of bespoke instrumentation for real world measurements and recognising the limitations of the original theoretical view revealed by experimental work.

## 1. Introduction

Lord Kelvin (the Irish-born William Thomson[1], 1824-1907), was Professor of Natural Philosophy at Glasgow University for over 50 years. He led a cascade of developments in fundamental physical understanding and their technological implementation (e.g. Whittaker, 1950), and is perhaps most famous for his eponymous temperature scale. However, his pioneering work in atmospheric electricity (e.g. Chalmers, 1967) developed in a short, intense period spanning 1859-1861 is little known and deserves wider appreciation (Simpson, 1929; McCartney, 2003; Harrison, 2002). This is both because of the ingenuity and originality apparent, but also because of the broader model his approach presents in applying physics to understanding the environment. Here the contributions Kelvin made to atmospheric electricity instrumentation and measurement are described, together with an assessment of their broader significance, both at the time and in today's context.

It is possible that Kelvin was inspired to work on atmospheric electricity by his interest in the growing telegraph network, in which he had commercial success through many patents (Trainer, 2004; 2007). The sensitive detection of small currents at the end of long telegraph lines, such as with his patented mirror galvanometer, may have shown atmospheric effects on the telegraph network. His detailed study of atmospheric electricity in turn motivated development of sensitive electrometers (Paterson et al, 1913; Simpson, 1929). Another motivation, apparent from his notebooks, was to test the application of potential theory, through atmospheric electrical measurements around flat and undulating surfaces. He also saw electrical data as providing a source of insight into atmospheric processes, encapsulated in his famous remark at the Royal Institution in 1860 that, "there can be no doubt but the electric indications, when sufficiently studied, will be found important additions to our means for prognosticating the weather" (REM[2] §287; Bennett and Harrison, 2007).

---

[1] This paper refers to Lord Kelvin, but lists references under the name Thomson.

Kelvin's work on atmospheric potentials provided standardised methods for their measurement. The existence of a positive potential in fair weather had been recognised since the late eighteenth century, and is now known to result from continued charging of the ionosphere, a conducting region in the upper atmosphere, by thunderstorms and shower clouds. The ionosphere is charged to a potential of 300kV more positive than the surface, which leads to a potential gradient of ~100Vm$^{-1}$ at the surface in fair weather. The weather was already established to cause electrical variations, such as in fogs (Bennett and Harrison 2007; Aplin et al, 2008). However, only relative measurements were possible, although retrospective analysis has since provided some calibrations for data from this early period (e.g. Harrison, 2009).

A recurrent theme in Kelvin's work is his surprise at the variability present in the atmospheric measurements he undertook. Almost certainly because of this, in 1859 he expressed the need for "incessant recording" of atmospheric parameters, by which he meant both continual measurement and constant recording of the data, in place of individual measurements that had to be made manually (Thomson, 1859a). Kelvin's instrumentation ultimately led to long-term measurements of atmospheric electricity that ran until the 1980s in the UK (Simpson, 1929; Harrison, 2003).

**2. Innovations in atmospheric potential measurements**

A fundamental atmospheric electrical measurement is to acquire the electric potential at a known height, from which the vertical potential gradient (PG) can be obtained. This requires that minimal distortion of the potential occurs from the measurement apparatus, or that a correction can be applied for the distortion. The potential itself is obtained by a probe that reaches equilibrium with the local potential of the air, and the probe's potential is measured with a high impedance electrometer. Pre-Kelvin, flame probes that generated local ionisation, or isolated conducting rods were used. Kelvin also used the flame probe, as well as developing two new sensors, described in section 2.1.

One of Kelvin's major innovations was to use electrochemical cells as an absolute voltage reference, which permitted the first calibrated PG measurements, at the Links golf course in Aberdeen at 8am on 14$^{th}$ September 1859, made with a flame probe,
> "The height of the match was 3 feet above the ground, and the observer at the electrometer lay on the ground to render the electrical influence of his own body on the match insensible. The result showed a difference of potentials between the earth (negative) and the air (positive) at the match equal to that of 115 elements of Daniel's battery." (Thomson, 1859a)

On the basis that that one zinc-copper Daniel cell generates an electromotive force of 1.08V, Bennett and Harrison (2007) calculated that the potential measured across 115 Daniel cells is 124.2V, which across a vertical distance of 3 feet (0.91m), gave a (vertical) PG of 137 Vm$^{-1}$.

Kelvin improved the sensing methods available for the atmospheric PG (section 2.1), applied the new technology of photography to data recording (section 2.2), and also designed a portable electrometer to allow roaming environmental measurements (section 2.3).

---
[2] "REM" refers to Thompson (1872), *Reprints of papers on electrostatics and magnetism,* with the paragraph number indicated.

## 2.1 Atmospheric Potential Gradient sensors

Lord Kelvin developed two PG sensors. The first, using mechanical exposure of a collecting electrode – the "winched can" sensor – determined the charge induced when an electrode was exposed to an electric field, making it, in some ways, an early precursor of the modern electric field mill. Its operation was cumbersome, and Kelvin devised a potential equaliser method soon afterwards. The "water dropper equaliser" operated on the principle of equalisation through the charge carried on water drops, with the great advantage that it could run continuously.

### 2.1.1 Winched can

Kelvin's winched can sensor was reported on 8$^{th}$ March 1859 by his friend James Joule at the Manchester Literary and Philosophical Society (Thomson, 1859b).

> "I have had an apparatus for Atmospheric Electricity put up on the roof of my lecture-room, and got a good trial of it yesterday, which proved most satisfactory."

The can sensor was developed from the equaliser rod used by Giovanni Battista Beccaria, who had measured fair weather atmospheric electricity in Turin towards the end of the eighteenth century, although Kelvin seemed unimpressed at Beccaria's need to sleep in the same room as the electrometer to obtain regular measurements (Thomson 1859a).

A similar instrument had been outlined by Kelvin in a notebook a few years earlier (Thomson, 1856; Figure 1). This was a hollow (possibly spherical) sensing electrode supported by an insulating glass rod, which was in turn attached to a metal mast. The sensing electrode was allowed to protrude about 60 cm above the roof of a building, in this case the Natural Philosophy Lecture Theatre at Glasgow University[3]. To make a measurement, a screening can, earthed using a long wire, was winched up to cover the electrode by a pulley. The earth connection was broken, the can lowered about 46 cm and its change in potential measured. Kelvin reported a substantial change in potential each time, "requiring more than one hundred degrees of torsion to bring it back to zero" which was a reference to his divided ring electrometer. The use of induction in the operation of the winched can sensor is reminiscent of a modern field mill instrument (MacGorman and Rust, 1998), although of course the covering and uncovering of a field mill's electrodes is now mechanised. Kelvin's notebooks contain a more detailed description of this instrument from about 1860 (Thomson, 1859e), transcribed in Appendix A.

Kelvin had previously described calibration of this type of electrometer against an electrochemical cell (REM §284), hence the torsion angles he describes can be understood to correspond to a change of at least 80V (Aplin, 2012). This is not unreasonable in a typical atmospheric PG of 100-200 $Vm^{-1}$, allowing for the likely geometrical distortion of the electric field by the building. Kelvin also corroborated the findings of Beccaria and others by showing that during "fair weather" the electrometer reading was positive, "but, when a negative cloud (natural, or of smoke) passes over, the indication is negative".[4]

Using this device, the extreme variability of the atmospheric PG, even under good conditions became apparent,

> "Even under a cloudless sky, without any sensible wind, the negative electrification of the surface of the earth, always found during serene weather, is constantly varying in degree." (Thomson, 1859a)

---

[3] The Natural Philosophy Lecture Theatre was at the old Glasgow University site on the High Street. The modern campus was not used until 1870.

[4] Here, Kelvin was observing smoke effects on PG. Kelvin electrograph measurements have been since used to deduce smoke pollution levels (Harrison and Aplin, 2002; 2003).

In the design of the winched can system Kelvin had separated the sensing aspect of the instrument (the electrode) from the recording aspect (the acquisition of potential by induction, and its measurement at a distant electrometer). This offered the possibility of making more frequent measurements. However, to explore PG variations more effectively, and with less manual effort, he invented a new instrument – the water dropper equaliser- which was capable of continuous measurements.

**2.1.2 Water dropper potential equaliser**
The first description of the Kelvin water dropper equaliser[5] was on 18th October 1859, again presented by Joule in Manchester (Thomson, 1859c). The description opened with,
> "I have a very simple 'domestic' apparatus by which I can observe atmospheric electricity in an easy way…"

The instrument comprised an insulated tank of water from which a continuous stream of water is allowed to flow (for example, out of a window), finally breaking into water drops. The water tank and an electrometer to measure the tank's potential would typically be installed on an upper floor of a building (Figure 2). At the stream to spray transition, droplets will polarise if their potential (which is that of the tank, via the connection provided by the water stream), differs from the local potential of the air. The effect of this polarisation at the moment of drop release is to cause charge transfer between the water stream and the air, which continues until the potential of the water stream equals the potential of the air at the stream-spray transition point. Because the tank is connected through the water stream to the spray formation point, the formation point potential can be measured more conveniently at the tank. If the height of the stream-spray transition point is also determined, the vertical potential gradient can be found.

The effect of continued charge transfer by droplets is that the water tank eventually acquires the same potential as that of the air at the stream-spray transition point. For example, if the tank potential is initially less than that of the air at the stream-spray transition point, there will a positive potential between the water stream and the air (i.e. a negative electric field). The negative field at the spray generation point implies a negative charge on the droplets generated, so that negative charge will be carried away from the water stream (and therefore the tank). The tank's potential will accordingly become more positive, until there is no local field between the air at the droplet generation point and the tank. Thus the tank acquires the same potential as the air at the spray generation point.

Electrostatic theory can be used to characterise the water dropper equaliser. When a drop leaves the stream it has a potential $V$ which results from both the potential of the stream at the position it is generated ($V_0$) – this is also the tank potential – and the charge $Q$ it is carrying,

$$V = V_0 + \frac{Q}{C_D} \qquad \text{Equation 1}$$

where $C_D$ is the capacitance of the drop (approximately $4\pi\varepsilon_0 a$ for a spherical droplet of radius $a$, where $\varepsilon_0$ is the permittivity of free space). The charge carried by the drop therefore depends on the difference in potential between the tank and the air, as

$$Q = C_D(V - V_0) \qquad \text{Equation 2}$$

When $V=V_0$, charge is no longer transferred and equalisation has occurred. If the capacitance of the entire water dropper system (the capacitance of the tank and stream) is $C$ and $n$ drops are generated per second, the rate of charge transfer is $nQ$, i.e. the current $i$ flowing is

$$i = 4\pi\varepsilon_0 an\Delta V \qquad \text{Equation 3}$$

for $\Delta V = V - V_0$.

---

[5] The water dropper potential equaliser should not be confused with the water dropper generator, a later invention of Kelvin (Thomson, 1867a). The generator uses electrostatic induction and positive feedback to produce high voltages. A video of a water dropper generator made at Oxford University is available at https://www.youtube.com/watch?v=nToCZuFXo70.

It is then possible to determine the effective resistance $R$ between the water dropper equaliser and the air as

$$R = \frac{\Delta V}{i} = \frac{1}{4\pi\varepsilon_0 an} \qquad \text{Equation 4}$$

No detailed information is available on the geometry or flow rates used, but for drops of radius 1mm falling at a rate of 1 s$^{-1}$, R~$10^{16}\Omega$, and for 100 μm radius drops, R~$10^{14}\Omega$. In each case this is a substantial resistance, providing a compromise between negligible loading effect on the atmospheric electric field, and an acceptable time response. The time response *RC*, of the dropper is 0.01-1s for the droplet sizes given above, consistent with Kelvin's own estimate of a time constant of less than 5s.

> "… any difference of potentials between the insulated conductor and the air at the place where the stream from the nozzle breaks into drops is done away with at the rate of five per cent, per half second, or even faster." (Thomson, 1859c)

The rapid response time of the water dropper equaliser reduced the demands on the insulator resistance, from which the charge only leaked away at 5% per minute, though Kelvin believed this could easily be improved to 5% per hour (Thomson, 1859c; 1859d). This time response, and the reduction in leakage by having the electrometer physically close to the water dropper equaliser, ensured a substantial improvement on previous instruments.

The water dropper equaliser remained Kelvin's preferred PG instrument, and was first used for a sustained period at his summer home on Arran (discussed in more detail in Section 3). In order to achieve his goal of "incessant recording", Kelvin needed to establish a data logging technique. This he did by applying the new technology of photography.

**2.2 Photographic recording**
Major advances in photography were made by Fox Talbot in the UK, and by the late 1850s the technology was well-established (Hannavy, 1997). Kelvin applied photographic techniques to scientific measurements by devising a technique where an electrometer's internal mirror was used to reflect a light (from a gas flame) onto a sheet of photographic paper, which was moved along by clockwork (Figure 2). As ever, Kelvin took care with calibration, by ensuring the position of the light spot for zero electric force, by reflecting another image of the gas flame onto the paper from a mirror attached to the electrometer case (REM §293). Photographic recording was naturally well-suited for use with an optical electrometer, and also provided minimal loading on the sensitive torsion balance. This combination of water-dropper and recorder became known as the Kelvin Electrograph, and its legacy is discussed in section 2.4. More information on the electrometers used is given in section 2.3.

Kelvin personally installed his electrograph at Kew Magnetic Observatory in January 1861 where it remained until May 1864 (Everett, 1868). The first known PG trace is shown in Figure 3. Relative measurements at Kew continued until 1898, when calibrations were applied to allow the absolute PG to be obtained (Chree, 1915). A water dropper equaliser system was used until 10$^{th}$ February 1932 when it was replaced with a radioactive probe (Harrison, 2006).

## 2.3 Portable measurements

Kelvin's work on electrometry appears to predate his atmospheric electricity measurements. He first presented his absolute electrometer in 1855 (Thomson, 1855), which was an improved version of his galvanometer[6], developed for the telegraphy industry and patented in 1858 (Trainer, 2004). Motivated by the need for better electrometers for "the observation of natural atmospheric electricity" (REM §303) and other applications, the divided ring electrometer was first described in 1856 (Thomson, 1867b). The divided ring electrometer operated by hanging a needle or rod, connected to the object to be measured, between two slightly separated halves of a horizontal ring, one charged up and one earthed. The charge on the needle disturbs the symmetry of this arrangement, and with the aid of a small mirror the needle's rotation can be detected optically. This system was readily adapted for photographic recording in the Kelvin Electrograph described in section 2.2. Kelvin subsequently developed the quadrant electrometer from the divided ring electrometer (Thomson, 1867b) which later became important for measurements in the UK's nascent electrical engineering industry (e.g. Wilson, 1898; Paterson et al, 1913).

A different principle of electrometry was to compare the force between two charged discs. This permitted absolute determination of the electromotive force, so the electrometers based on this principle were known as absolute electrometers (Thomson, 1867b). Kelvin developed a portable electrometer during his productive "electrostatic summer" on Arran in 1859, which also operated on the "attracted disc" principle. It consisted of two concentric, vertically separated Leyden jars with a small bar of aluminium, the "index", suspended between them on fine platinum wire (Figure 4, Figure 5). A conductor was connected to the body to be tested, and its potential caused vertical repulsion or attraction of the index. The upper Leyden jar was removable and rotatable, and a micrometer allowed measurement of the torsion needed to bring the index back to a fixed position. The difference between an earth and "air" reading gives the potential, which is proportional to the square root of the torsion angle (REM §277). A key aspect of Kelvin's approach can again be seen: he carefully referenced the instrument's response to electrochemical cells, making it possible to interpret his measurements today.

Kelvin appears to have been inspired to produce the portable electrometer by the need to make roaming atmospheric PG measurements, with a flame probe. Though Kelvin claimed the apparatus was, "easy to carry … to the place of observation, even if up a rugged hill side, with little risk of accident" (REM §277), it might today be described as "movable" rather than "portable", Figure 6. He subsequently realised the portable electrometer's broader utility, and collaborated with an instrument maker, James White, for commercialisation of the technology (Trainer, 2004)[7]. Measurements made with the portable electrometer are discussed further in section 3.

## 2.4 Kelvin's legacy to atmospheric electricity instrumentation

In summary, Kelvin was responsible for several innovations in his brief period working on atmospheric electricity.

1. He recognised the need for continuous measurements.
2. He used electrochemical cells to provide calibrated measurements.
3. He developed the winched can sensor, for more convenient and rapid measurements.
4. He developed the water dropper equaliser sensor, which could operate continuously.

---

[6] Such electrometers measured potential difference through electrostatic force effects, whereas galvanometer measurements of current could also be used (Thomson, 1867b).
[7] One anecdote describes how Lord Kelvin asked another Glasgow professor if he could recommend an instrument maker who might be good at producing electrometers commercially. Mr White, a skilled microscope maker was recommended, but Kelvin's colleague subsequently found that he was unable to get hold of microscopes any more, since Mr White's shop had rapidly become full of Kelvin's electrometers (Wilson, 1913).

5. He implemented photographic recording for use with the water dropper equaliser.
6. He made sensitive electrical measurements possible outside a laboratory environment for the first time through a portable instrument.

The water dropper equaliser technique, in combination with photographic recording first used at Kew, has left a long legacy to atmospheric science, summarised in Table 3. Eskdalemuir Observatory, which was established as a magnetically quiet alternative location to Kew, commenced regular, standardised PG measurements with water dropper equalisers in 1908. This continuity of measurement did much to establish regular atmospheric electrical monitoring by the Met Office, which continued in the UK until the early 1980s (Harrison, 2003).

Water dropper equaliser technology was adopted internationally, and also deployed for early measurements above the surface. Tuma (1899) used a triple water dropper equaliser system on a balloon ascent to ~4km, described in Nicoll (2012), and water dropper equalisers were also installed at the top of the Eiffel Tower in 1889 (Harrison and Aplin, 2003). Water dropper equalisers are still used for routine observation at the Kakioka monitoring site in Japan, with which PG changes were observed following the radioactivity release at Fukushima (Takeda et al, 2011).

**3. Atmospheric electricity measurements**
Lord Kelvin's letters and publications indicate that he was at his most active in experimental atmospheric electricity work from 1859 to 1861 when he installed the electrograph at Kew Observatory. His 1859 measurements were made exclusively in Scotland, starting at Glasgow University with the winched can described in section 2.1. A letter to Kelvin from his technical assistant Macfarlane establishes that the Glasgow measurements continued until at least July 9$^{th}$ 1859. There is also the evidence of measurements made at Aberdeen, as described in Section 2. A summary of Kelvin's measurements and activities in atmospheric electricity in Scotland in 1859 is given in Table 1.

**3.1 Kelvin's work on Arran in 1859**
Kelvin had enjoyed prolonged summer holidays at a range of rented houses on Arran since childhood (King, 1909) a practice he continued in adult life. Detailed reports exist of a series of atmospheric electricity experiments he undertook on Arran in summer 1859. During 1859 Kelvin stayed at houses called "The White Block" in the Invercloy part of Brodick and "Kilmichael" which is at Glencloy, inland from Brodick (Inglis, 1930; Thomson, 1859e)[8].

Kelvin's notebooks (Thomson, 1859e) indicate that, in late June 1859, he attempted to apply electrostatic theory to calculate distortion of the surface potential gradient by the topography of islands and mountains (Table 1). The excerpt in Figure 7 shows a sketch of an island shaped not unlike Arran. Other sketches in the notebook (Thomson, 1859e) show mountains reminiscent of Arran's highest peak, Goatfell. His familiarity with Arran may have influenced these general considerations or he may even have been trying to simulate particular sites. It is not clear from the notebooks whether Kelvin was anticipating, or quantitatively accounting for likely electrostatic distortions, but as but the first evidence for work on the portable electrometer is also in July 1859, this could have been motivated by Kelvin's intention to test his potential distortion considerations experimentally.

---

[8] "The White Block" is now called Castle View. Kelvin's notebook entry for 20$^{th}$ June 1859 is referenced "Invercloy", suggesting that he was staying at Castle View at this time. The notebook entry for 20$^{th}$ September 1859 is marked "Kilmichael", which is now a hotel. It is unclear how much time Kelvin spent at Kilmichael in 1859, but he had visited the house regularly as a guest of its owners, Captain and Mrs Fullarton, as a child (King, 1909). The grounds of Kilmichael also contained an astronomical observatory, established by a Dr Robertson-Fullarton (MacBride, 1910). The remains of the telescope have been retained by a local farmer.

Kelvin installed both winched can and water dropper equaliser sensors at his Arran houses in 1859 (Thomson, 1859c; 1860), but, as he only ever refers to one sensor at once, it appears unlikely that both were used simultaneously. Inglis (1930) refers to atmospheric electrical measurements at The White Block; the description of local people being alarmed by Kelvin drawing sparks from the roof strongly implies that the winched can sensor was used there. Kelvin's letters from August 1859 (Table 1) also imply a winched can sensor at The White Block (Invercloy). A subsequent description from October 1859 explains that, by then, a water dropper equaliser had been installed (Thomson, 1859c).

Within four months of his theoretical work on the electrostatic distortion of mountains, Kelvin achieved his goal of measuring the PG (with a flame probe and the portable electrometer) at the beach and on Goatfell. Typical fair weather measurements on Arran, expressed in modern units, were 80-158 $Vm^{-1}$, consistent with other clean air Scottish sites (Thomson, 1859c; Aplin, 2012). Kelvin observed that, "even in fair weather, the intensity of the electric force in the air near the earth's surface is perpetually fluctuating…" (REM §282) and pointed out that at Arran, the PG always increased substantially when the wind was from the east or north-east. Aplin (2012) has interpreted these enhanced PGs as the reduction in air conductivity by smoke pollution from the industrial western Scottish mainland. At the time however, Kelvin was clearly as surprised by the PG variability as he had been for Glasgow measurements.

Despite his measurement successes, it is not clear that Kelvin was able to make sense of the PG distortion caused by Goatfell, as no further reference to this work on potential theory has been found in his notebooks. Kelvin's simultaneous measurements at Brodick and Goatfell appear to have confirmed that the PG varies between different locations, rather than being of any practical use in confirming or refuting his theoretical predictions. Kelvin's conclusions on such a combined study ultimately appeared much later:
> "Hence, if we find observations made simultaneously by two electrometers in neighbouring positions, in a mountainous country, to bear always the same mutual proportion, we may not be able to draw any inference as to electrified air; but if, on the contrary, we find their proportion varying, we may be perfectly certain that there are varying electrified masses of air or cloud not far off." (REM §262)

**3.2 Measurements during the Carrington event**
Kelvin's PG measurements on Arran occurred around the time of the "Carrington event". This was the strongest solar flare ever recorded, noticed by Richard Carrington at his solar observatory in Redhill, Surrey, UK at 1115 GMT on 1$^{st}$ September 1859. The Carrington event caused widespread aurorae, even in the tropics, and serious disturbances to the telegraph network (e.g. Clark, 2009; Cliver, 2006). Retrospective analysis indicates that the event comprised two solar flares, with a smaller emission on 27$^{th}$ August preceding the major event on 1$^{st}$ September. Solar proton fluxes did not return to their usual levels until about 8$^{th}$ September (Smart et al, 2006). There is evidence that solar activity can cause changes in atmospheric electricity through the global electric circuit (e.g. Olson, 1971; Mironova et al, 2013). As the 1$^{st}$ September 1859 solar flare was almost certainly associated with a "ground level event", when energetic solar protons reach the ground (Smart et al, 2006), an effect on surface atmospheric electricity through enhanced ionisation can also be expected.

Kelvin's letters, notebooks and papers summarise his activities during the Carrington solar storm. Table 1 shows that Kelvin was very pleased with the progress with his portable electrometer on 6$^{th}$ August, but then, during late August, he wrote to his friend Joule in Manchester asking for help. Joule's response on 5$^{th}$ September acknowledges Kelvin's request:

> "You say you would like some help with your portable electrometer at which I do not wonder. Such an apparatus appears to me to be the most difficult thing imaginable to render portable. " (Joule, letter to Kelvin, 5$^{th}$ September 1859)

Unfortunately Kelvin's letter is lost, and consequently the nature of the help required is not known, but as Kelvin could obtain technical support from Glasgow University, it seems likely that the help requested would be concerned with scientific, rather than technical difficulties.

The question of when Kelvin's lost letter was written is now briefly considered further, by examining the dates of Joule's previous and subsequent letters. Joule wrote to Kelvin on 22$^{nd}$ August, and again, as mentioned, on 5$^{th}$ September. Assuming the post from Manchester to Arran would have taken at least three days by rail and steamer[9], Kelvin's reply to Joule would have been written in the nine days between 25$^{th}$ August and 2$^{nd}$ September. Solar activity was considerably enhanced during this window, with aurorae frequently observed. It is therefore probable that Kelvin was attempting to measure the PG during the disturbed Carrington flare period, and wrote to Joule for help with his unusually problematic electrometer (Table **2**). A week after the Carrington flare, Joule and Kelvin successfully obtained PG measurements in Aberdeen (section 2). The electrometer had been functioning reliably earlier in August (Table 1) and results are also available from October (Aplin, 2012). In summary, as Table **2** shows, Kelvin's request for help with the portable electrometer was almost certainly written during strongly disturbed solar conditions, which cannot be excluded as a source of the anomalous electrical measurements encountered.

**4. Discussion**

Kelvin's Scottish atmospheric electricity work exhibits a combination of theoretical expectations and experimental results. This presents an example of the early Victorian scientific method as outlined by Whewell, through which scientists try to unravel the "book of nature" provided by a Creator. Within this conceptual framework the full text of the book is of course inaccessible, but the application of mathematics can give context (Gower, 1999). Kelvin was probably one of the first to use this quantitative approach within the geosciences.

More specifically, an unexpected finding of the Glasgow measurements was that the atmospheric electric field was a highly variable quantity on different timescales. In making the subsequent measurements on Arran, Kelvin may have been acknowledging the local effects of the city and was seeking a more quiescent natural measurement environment after the effects of topographic distortion had been considered. Even so, the variability persisted. It is now known that, as well as changes in the weather, as recognised by Kelvin, this could arise from effects of local pollution or even, possibly, solar activity. The variability consistently present in the experimental measurements therefore challenged Kelvin's theoretical view of the electrostatic environment. Reconciliation of the simplifications necessary for theoretical description with the reality of experiments undertaken in the natural world is a characteristic of environmental physics work in general. The atmospheric electricity work of Kelvin can therefore also be seen as a pioneering example of environmental physics.

Kelvin's view of atmospheric electricity was based on the then prevalent electrostatic view, before the continuous replenishment of atmospheric ionisation was established at the end of the nineteenth century (Simpson, 1906). This provides some insight into why Kelvin saw a need for incessant recording of electrical parameters, and remarked on changes in the local weather associated with atmospheric electrical changes, as they amounted to perturbations on what was expected to be a settled background state. In this, his geophysical viewpoint can be seen to be similar to that concerning the age of the earth (Thomson, 1863), which was derived from considerations of a system cooling, also to a steady state. Whilst in no sense a creationist standpoint, the Victorian view of an inherited earth system, informed by the "book of nature" approach mentioned above, may have contributed to this expectation of steady electrostatic conditions[10]. Subsequent independent work on ionisation of air by Elster and Geitel (1900) in Germany and CTR Wilson (1901) in Scotland established the need for atmospheric current flow, and with it, ultimately the concept of a varying, dynamic, terrestrial atmospheric electricity system.

---

[9] Bradshaw's railway guide for September 1859 indicates that the rail journey from Glasgow to Manchester took approximately 12 hours. Clearly, letters between Manchester and Arran would also have to travel between Arran and the mainland by steamer.

In terms of atmospheric electricity advances, Kelvin's work provides the first calibrated PG measurements. The experiment on 10$^{th}$-11$^{th}$ October on Arran also represents the first coordinated multi-site measurement of PG and the experimental techniques were later used for routine PG measurements at Kew Observatory. Water-dropper equalisers were subsequently established at other observatories in the early twentieth century, producing a long data set of PG and other atmospheric electrical quantities (Harrison, 2003; Takeda et al, 2011). The historical atmospheric electricity data sets from the UK observing network have proved useful in studying long-term atmospheric change, particularly that of air pollution (Harrison, 2006) and responses to solar events (Harrison and Usoskin, 2010). It remains possible that Kelvin himself may have inadvertently detected a surface atmospheric electrical response to the major solar event in the late summer of 1859. This would be ironic in the context of Kelvin's firm view that solar activity and magnetic storms were unrelated, and that the sun was not capable of inducing electromagnetic changes detectable at Earth (Thomson, 1892).

Notwithstanding the explanatory framework in which they were originally undertaken, Kelvin's well-calibrated atmospheric electricity measurements were so fully described and implemented that the scientific methodology, instrumentation technology development and observational findings remain completely relevant today.

**Acknowledgements**
John Riddick (formerly Eskdalemuir Observatory) provided historical documents prepared by staff there. Matthew Trainer (Glasgow University) assisted with the work on the portable electrometer and prepared Figure 5. Aurora observations were obtained from the Armagh Observatory archive (http://climate.arm.ac.uk).

7. **References**

Aplin, K.L. (2012), Smoke emissions from industrial western Scotland in 1859 inferred from Lord Kelvin's atmospheric electricity measurements, *Atmos. Env.*, **50**, 373-376 doi:10.1016/j.atmosenv.2011.12.053

Aplin K.L., Harrison R.G. and Rycroft M.J. (2008), Investigating Earth's atmospheric electricity: a role model for planetary studies, *Space Sci. Rev.* **137**, 1-4, 11-27 doi: 10.1007/s11214-008-9372-x

Bennett A.J., Harrison R.G. (2007), Atmospheric electricity in different weather conditions *Weather* **62**, 10, 277-283

Chalmers J.A. (1967) *Atmospheric electricity*, 2$^{nd}$ edition, Pergamon Press, Oxford, UK

Chauveau, A.-B. (1893), Sur la variation diurne de le électricité atmosphérique, observeé au voisinage du sommet de la tour Eiffel, Comptes Rendus **67**, 1069–1072.

Chree C. (1915), Atmospheric electricity potential gradient at Kew Observatory, 1898 to 1912, *Phil. Trans. Roy. Soc. Lond*. **215**, 133-159

Clark S.G. (2009), *The Sun Kings*, 2nd edition, Princeton University Press, USA

Cliver E.W. (2006), The 1859 space weather event: Then and now, *Adv. Space Res*. 38, 119–129
---

[10] Lord Kelvin was an Elder of the Church of Scotland in his later life (Thompson, 1910)

## Appendix A: "On contact electricity": a description of the winched can sensor from Kelvin's notebook

Here, Kelvin's description of the winched can sensor has been transcribed from Notebook 35. Repeated drafts of almost identical sentences, and a lengthy footnote have been omitted. The notebook entry is not dated, but the references in the footnote indicate it was written in or after 1860. The first page of this description is shown in Figure 8. The "second drawing" referred to in the excerpt is not in the notebook.

> On contact Electricity
> In one of the methods which I have employed for observing atmospheric electricity a metal conductor, cylindrical with flat ends, about [blank] inches long, and [blank] diameter, which will be called the collecting conductor, was insulated on the top of an iron rod, at a height of three or four feet above the roof of the observatory. A cylindrical metal can with its mouth up, was arranged to ride on this iron rod passing through a hole in its bottom, & maintained by counterpoises in any position in which the observer leaves it after moving up or down by means of a rod passing through a collar in the roof into the observing room below. A fine platinum wire, which, after

Beccaria, will be called the <u>deferent wire</u>, attached to a point in the bottom corner of the insulated conductor and passing through a wide hole in the moveable can, and another in the roof.

The movable can was kept constantly in metallic communication, by means of the iron rod, & with the metal case enclosing the electrometer case. When an observation is to be made, the can must be first in its highest position, and when it is there, the collecting conductor, deferent wire & principal electrode of the electrometer are, by means of a convenient metallic arc, made to communicate with the disinsulated metallic system connecting with the electrometer case, and the zero, or earth reading, of the electrometer is taken. The disinsulating metallic arc is then removed, so as to leave the metallic system of which the collecting conductor is part, insulated and the can is drawn down into the position shown in the second drawings. The electrometer immediately indicates the proper influence of atmospheric electricity for the observation of which the apparatus is constructed.

If the insulation were perfect the observation might be continued indefinitely, and atmospheric influence, always varying as it is, would be always correctly indicated in the electrometer, without any renewal of the operation first described; the principle being that the insulated trace, having been once perfectly discharged, being disinsulated when the can is up can never afterward, if perfectly insulated, acquire any absolute charge.

8. Figures and Tables

| Period in 1859 | Location / activity | Evidence | Inference |
|---|---|---|---|
| January-March | • In Glasgow, measuring PG at the Natural Philosophy Lecture Theatre on the High Street | • Thomson, 1859b | |
| June | • In Brodick, staying at Invercloy | • Thomson, 1859e<br>• Working on the geometrical distortion of the atmospheric electric field by mountains | |
| July | • In Brodick, staying at Invercloy<br>• Measurements in Glasgow until at least 9th July<br>• Met assistant Macfarlane on Wednesday 13th July? | • Envelope for Letter to Thomson, postmarked 1st July 1859 (Glasgow MS Kelvin T162)<br>• Letter received from Macfarlane, dated 9th July 1859 (Cambridge M11). Describes testing of an electrometer and ends letter with, "… which you will see on Wednesday"<br>• Envelope for Letter to Thomson, postmarked 30th July 1859 (Glasgow MS Kelvin T163) | Could have been given the portable electrometer to work on by Macfarlane when they met. |
| August | • Working on atmospheric electricity at Invercloy.<br>• Winched can sensor on roof of house | • Letter to John Pringle Nichol, 5th August 1859 (Cambridge N35), "An atmospheric electrometer has kept me from putting pen to paper these last few days."<br>• Letter to James Thomson, 6th August 1859, "I am at work only on atmospheric electricity at present…My atmospheric electrometer is in good action. Three days ago I had splendid effects during showers … the conductor only clamped about 8 feet above the roof of the building" (Glasgow MS Gen 1752/3/1/T/70)<br>• Letter to from Jenkin 9th August 1859 (Glasgow MS Gen 1752/5/1/J/3)<br>• Letter from Joule, Mon 22nd | Only the winched can sensor could have been mounted on a roof |
| | | August 1859, saying Joule will visit Arran (Glasgow MS Kelvin J150) | |
| 1st-13th September | • Working on atmospheric electricity at Invercloy<br>• May have met Joule on Saturday 10th September | • Letter from Joule, Mon 5th September 1859, saying he is visiting "on Saturday" and acknowledges request for help with portable electrometer (Glasgow MS Kelvin J151) | Assuming the postal service took 3 days, then Kelvin must have replied to Joule between 25th August - 2nd September, indicating that he was testing the electrometer during this period |
| 14th September | • Measurements at Aberdeen, with Joule | • Thomson, 1859a | |
| Early October | • At Invercloy, working on portable electrometer. | • Letter to unknown correspondent planning measurements on Goatfell, 3rd October 1859 (Glasgow MS Gen 526/2)<br>• Envelope addressed to Prof Thomson from J.D. Forbes, Glencloy, Arran, 6th October 1859 (Glasgow MS Kelvin T164) | |
| 10th-11th October | • Measurements on Goatfell, Brodick Beach and simultaneously at house with water-dropper | • Thomson (dates of experiment reported incorrect in 1859c, corrected to October in 1859d)<br>• Aplin (2012) | |
| 18th October | • At Invercloy, testing water dropper equaliser | • Thomson, 1859c | |

Table 1 Kelvin's 1859 Scottish PG measurements. Letter references refer to the local classification at the library where the letter is held (Cambridge or Glasgow University Library).

| Date in 1859 | Joule | Kelvin | Solar activity |
|---|---|---|---|
| 22nd August | Letter from Joule to Kelvin | working on electrometer | faint aurora at Armagh to north-west |
| 23rd | *Postal transit of Joule's letter* | | |
| 24th | | | |
| 25th | | | faint aurora at Armagh to north-east |
| 26th | | | |
| 27th | | | |
| 28th | | Kelvin writes to Joule asking for help with the electrometer implying it was no longer working well (letter lost) | solar proton event; faint aurora at Armagh |
| 29th | | | powerful aurora at Armagh seen despite cloud |
| 30th | | | |
| 31st | | | |
| 1st September | | | Coronal Mass Ejection arrival at Earth; Carrington observes solar flare by eye at 1115 |
| 2nd | *Postal transit of Kelvin's lost reply* | | bright aurora from north-east to west south west at Armagh; |
| 3rd | | | strong aurora at Armagh |
| 4th | | | strong aurora at Armagh |
| 5th | Letter from Joule to Kelvin | | |

Table 2 Summary of Kelvin's interaction with Joule during the Carrington flare period

| Years | Location | Notes | Reference |
|---|---|---|---|
| 1861-1932 | Kew Observatory, London | Standardised observations commenced in 1898 | Everett (1868), Harrison (2006) |
| 1893, 1896-1898 | Eiffel Tower, Paris | Located on top | Chaveau (1893), Harrison and Aplin (2003) |
| 1899 | Austria | Balloon measurements | Tuma (1899), Nicoll (2012) |
| 1908-1930s | Eskdalemuir Observatory, Scotland | Systematic hourly calibrated PG measurements | Harrison (2003) |
| 1962-present | Kakioka Magnetic Observatory, Japan | Systematic hourly measurements used to observe effect of radioactive fallout from Fukushima | Takeda et al (2011) |

Table 3 Use of the water-dropper equaliser in atmospheric electricity measurements.

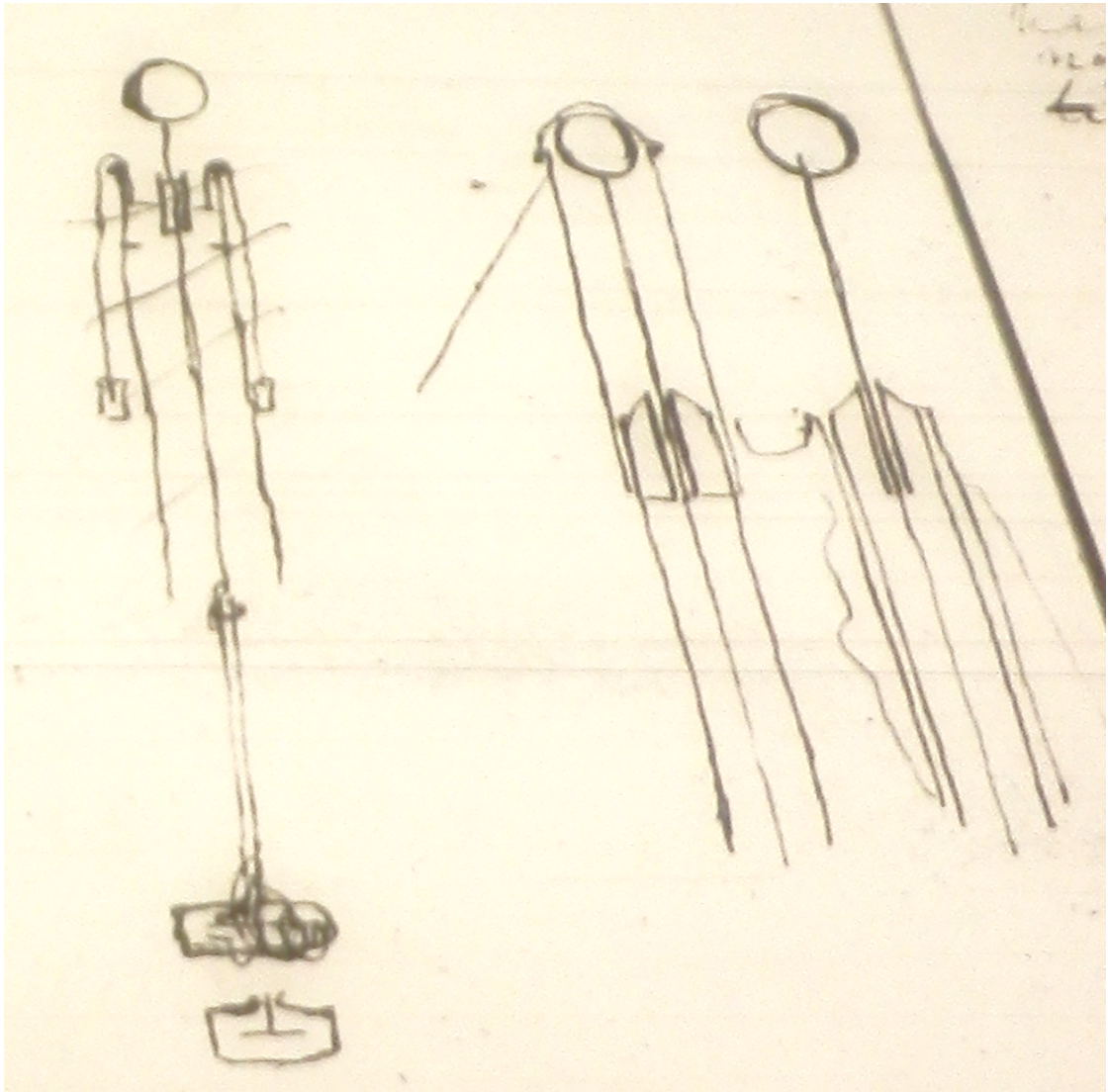

**Figure 1 First sketches of an instrument for measuring potential gradient, from Thomson (1856). The sketch is marked "Observations of atmospheric electricity" and on the same page is a sketch for an electroscope to be used with the instrument. The sketches that are not crossed out may indicate a method for exposing and enclosing the ball, and looking for the induced change, similar to the principle of operation of the winched can sensor and its 20$^{th}$ century version, the field mill.**

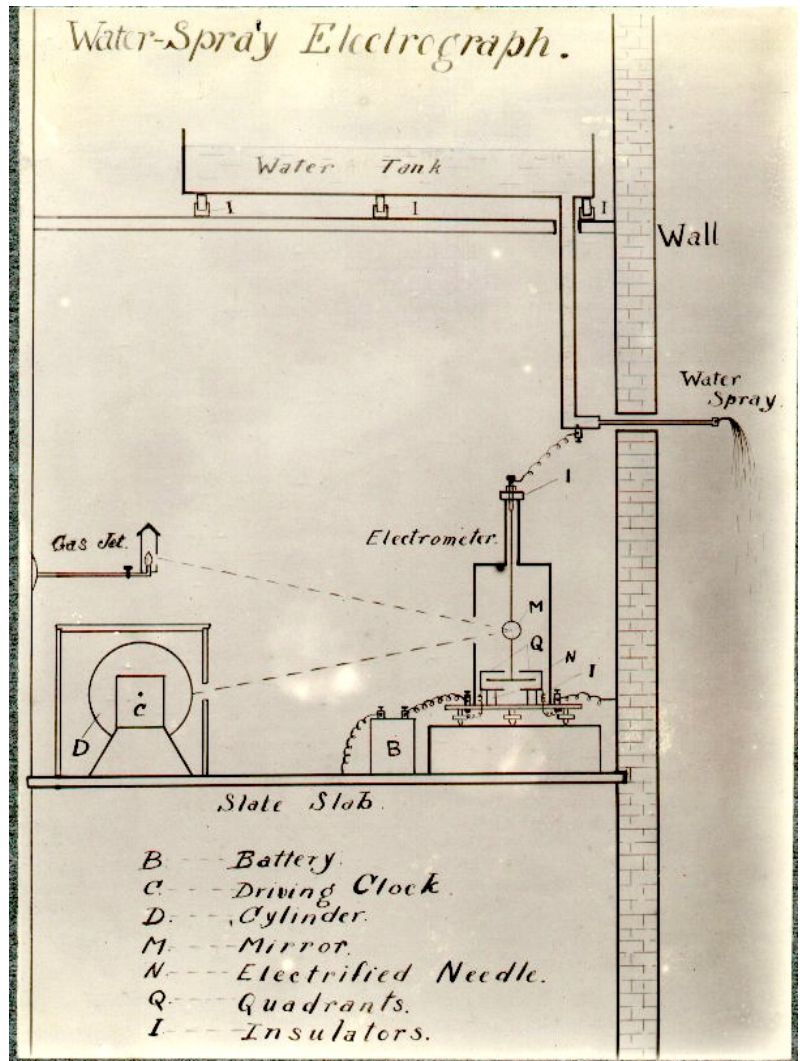

**Figure 2 Description (right) and diagram (left) of the water-dropper apparatus, showing photographic recording. Taken from Gendle (1912).**

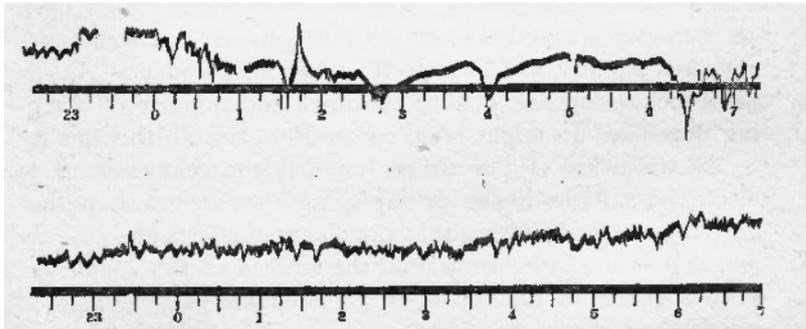

**Figure 3 Photographically-recorded PG traces from Kew Observatory, 28th and 29th April 1861 (from REM §292)**

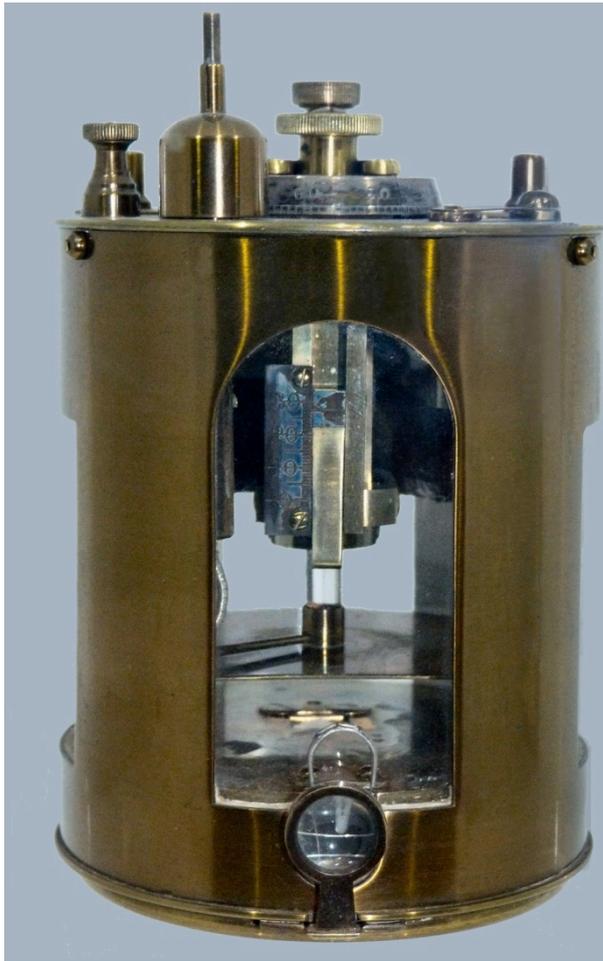 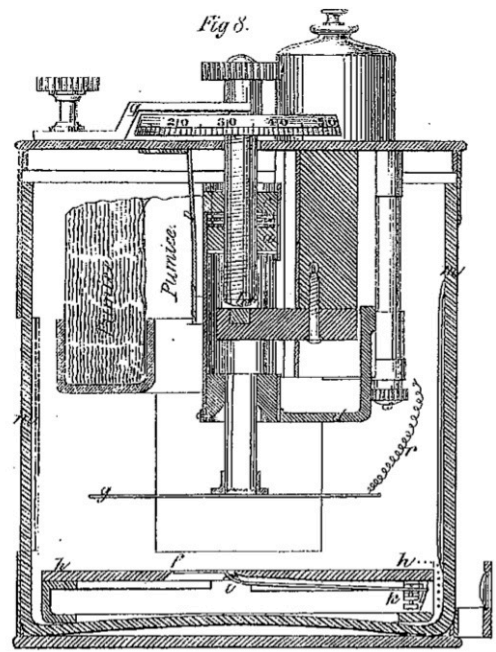

**Figure 4 (left) photograph of the portable electrometer, courtesy of Glasgow University Library (right) technical drawing, taken from Thomson (1872).**

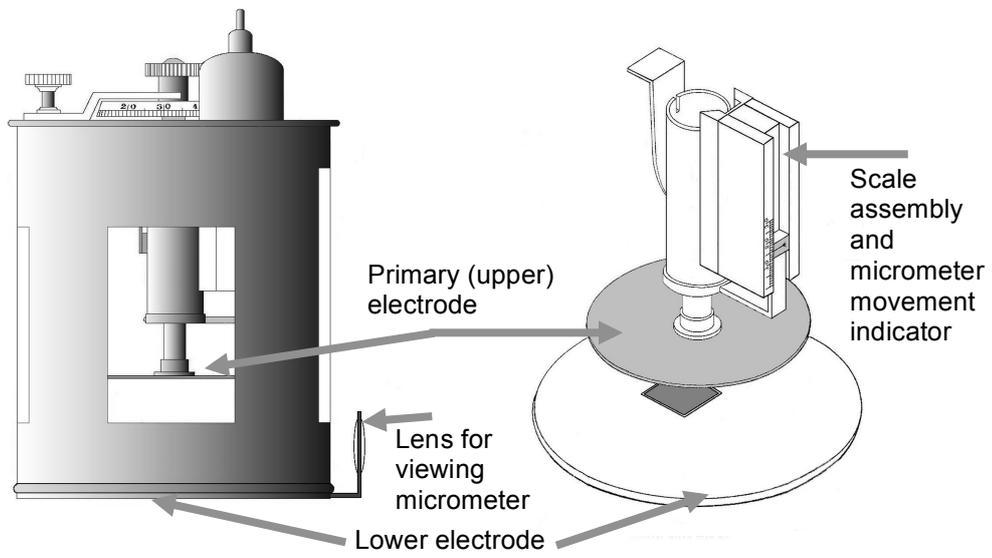

Figure 5 Simplified diagram of the Kelvin portable electrometer, showing outer casing (left) and (right) major internal components.

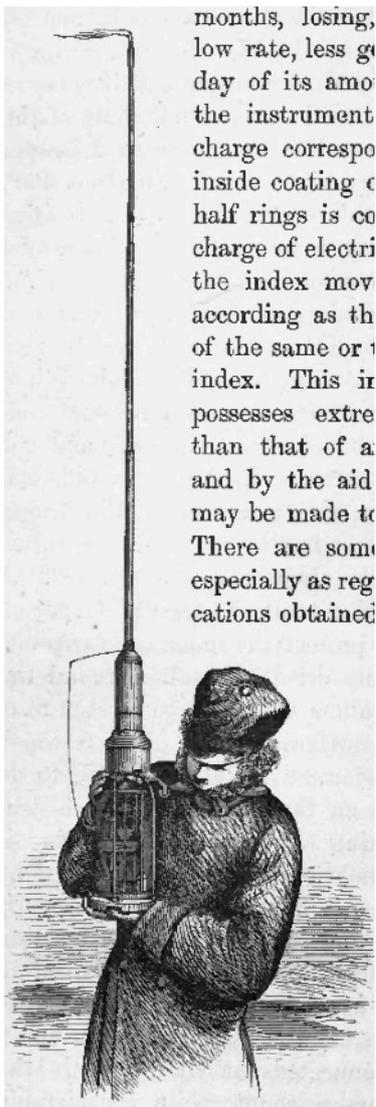

**Figure 6 Drawing of a Kelvin portable electrometer, attached to a flame probe and carried by an assistant. Note the rural location, probably Arran. From REM §263.**

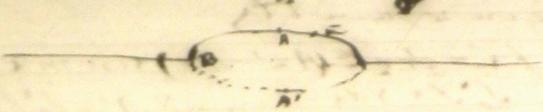

**Figure 7** Excerpt from Notebook 35 (Thomson, 1859e), dated June 21st 1859 and written at Invercloy (top left), showing Kelvin's calculations and sketches of the electric field distortion due to mountains and islands.

## On Contact Electricity

In one of the methods which I have employed for observing atmospheric electricity a metal conductor, cylindrical with flat ends, about _____ inches long and _____ diameter, which will be called the collecting conductor, was insulated on the top of an iron rod, at a height of three or four feet above the roof of the observatory. A cylindrical metal can, with its mouth up, was arranged to slide on this iron rod passing through

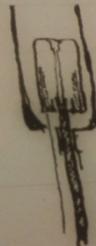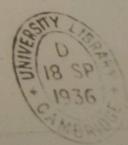

**Figure 8 Page fron Notebook 35 describing the winched can sensor, with a sketch of the instrument.**